\documentclass[fleqn,usenatbib]{mnras}
\usepackage{float}

\usepackage{newtxtext,newtxmath}

\usepackage[T1]{fontenc}

\DeclareRobustCommand{\VAN}[3]{#2}
\let\VANthebibliography\thebibliography
\def\thebibliography{\DeclareRobustCommand{\VAN}[3]{##3}\VANthebibliography}

\usepackage{graphicx}	
\usepackage{amsmath}	
\usepackage{pdflscape}
\usepackage{subcaption}
\title[Constraining dark matter signal from the TGSS]{Stringent constraint on the radio signal from dark matter annihilation in dwarf spheroidal galaxies using the TGSS}

\author[Basu et al.]{Arghyadeep Basu,$^{1}$\thanks{E-mail: arghyadeepbasu.97@gmail.com} Nirupam Roy,$^{2}$ 
Samir Choudhuri,$^{3}$ Kanan K. Datta$^{1}$ \newauthor and Debajyoti Sarkar$^{1}$\\
$^{1}$Department of Physics, Presidency University, Kolkata 700073, India\\
$^{2}$Department of Physics, Indian Institute of Science, Bangalore 560012, India\\
$^{3}$Astronomy Unit, Queen Mary University of London, Mile End Road, London E1 4NS, United Kingdom
}

\date{Accepted XXX. Received YYY; in original form ZZZ}

\pubyear{2020}

\begin{document}
\label{firstpage}
\pagerange{\pageref{firstpage}--\pageref{lastpage}}
\maketitle

\begin{abstract}
Weakly Interacting Massive Particles (WIMPs) are considered to be one of the favoured dark matter candidates. 
Searching for any detectable signal due to the annihilation and decay of WIMPs over the entire electromagnetic spectrum has become a matter of interest for the last few decades. WIMP annihilation to Standard Model particles gives rise to a possibility of detection of this signal at low radio frequencies via synchrotron radiation. Dwarf Spheroidal Galaxies (dSphs) are expected to contain a huge amount of dark matter which makes them promising targets to search for such large scale diffuse radio emission. In this work, we present a stacking analysis of $23$ dSph galaxies observed at low frequency ($147.5$ MHz) as part of the TIFR-GMRT Sky Survey (TGSS). The non-detection of any signal from these stacking exercises put very tight constraints on the dark matter parameters. The best limit comes from the novel method of stacking after scaling the radio images of the individual dSph galaxy fields after scaling them by the respective half-light radius. The constraint on the thermally averaged cross-section is below the thermal relic cross-section value over a range of WIMP mass for reasonable choices of relevant astrophysical parameters. Such analysis using future deeper observation of individual targets as well as stacking can potentially reveal more about the WIMP dark matter properties.  
\end{abstract}




\section{Introduction}
\label{section:1}

The hierarchical structure formation model predicts existence of numerous low mass satellite galaxies gravitationally bound to larger, high mass galaxies. For the Milky Way (and the local group), a large fraction of these observed satellite galaxies are dwarf spheroidal (dSPh) galaxies. The dSph galaxies, very similar to the globular cluster population, are gravitationally bound, diffuse, low surface brightness and low luminosity systems \citep{massscale}, with low metallicity and old stellar population \citep{McConnachie}, and having negligible gas and dust content, as well as very little or no recent star formation. However, unlike the globular clusters, they have very high dynamical mass to light ratio \citep{galaxy}, indicating that they are dark matter (DM) dominated systems \citep[e.g.][]{massscale,Louis}. The presence of dark matter can also be inferred from the kinematics studies of dSph galaxies \citep[e.g.][]{bosma,Rubin}. Based on a variety of observations, the dSph galaxies are thought to be the most dark matter dominated systems, and hence are considered to be among the most promising targets for the indirect detection of dark matter \citep{Louis2}.

\setlength{\tabcolsep}{5pt}
 \begin{table}
    \begin{tabular}{lcccc}
      \hline
      Name           & R.A.          & Dec.              & $r_{h}$($\arcmin$)     &D ($kpc$)   \\
      \hline
      Aquarius       & $20h46m52s$ & $-12\degr50'33"$ & $1.47\pm0.04$          &$1072\pm39$ \\
      Bootes I       & $14h00m16s$ & $+14\degr30'00"$ & $12.6\pm1$             &$66\pm2$    \\
      Bootes II      & $13h58m00s$ & $+12\degr51'00"$ & $4.2\pm1.4$            &$42\pm1$    \\
      Carina         & $06h41m37s$ & $-50\degr57'58"$ & $8.2\pm1.2$            &$105\pm6$   \\
      Cetus          & $00h26m11s$ & $-11\degr02'40"$ & $3.2\pm0.1$            &$755\pm24$  \\
      Columba I      & $05h31m26s$ & $-28\degr01'48"$ & $1.9\pm0.5$            &$182\pm18$  \\
      Draco          & $17h20m12s$ & $+57\degr54'55"$ & $10.0\pm0.3$           &$76\pm6$    \\
      Fornax         & $02h39m59s$ & $-34\degr26'57"$ & $16.6\pm1.2$           &$147\pm12$  \\
      Grus I         & $22h56m42s$ & $-50\degr09'48"$ & $1.77^{+0.85}_{-0.39}$ &$127\pm6$   \\
      Grus II        & $22h04m05s$ & $-46\degr26'24"$ & $6.0^{+0.9}_{-0.5}$    &$53\pm5$    \\
      Hercules       & $16h31m02s$ & $+12\degr47'30"$ & $8.6^{+1.8}_{-1.1}$    &$132\pm12$  \\
      Hydra II       & $12h21m42s$ & $-31\degr59'07"$ & $1.7\pm0.3$            &$134\pm10$  \\
      Leo I          & $10h08m28s$ & $+12\degr18'23"$ & $3.4\pm0.3$            &$254\pm15$  \\
      Leo II         & $11h13m29s$ & $+22\degr09'06"$ & $2.6\pm0.6$            &$233\pm14$  \\
      Leo IV         & $11h32m57s$ & $-00\degr32'00"$ & $4.6\pm0.8$            &$154\pm6$   \\
      Leo V          & $11h31m10s$ & $+02\degr13'12"$ & $2.6\pm0.6$            &$178\pm10$  \\
      Phoenix        & $01h51m06s$ & $-44\degr26'41"$ & $3.76$                 &$415\pm19$  \\
      Sagittarius II & $19h52m40s$ & $-22\degr04'05"$ & $2.0\pm0.4$            &$67\pm5$          \\
      Sculptor       & $01h00m09s$ & $-33\degr42'33"$ & $11.3\pm1.6$           &$86\pm6$    \\
      Segue II       & $02h19m16s$ & $+20\degr10'31"$ & $3.4\pm0.2$            &$35\pm2$    \\
      Ursa Major II  & $08h51m30s$ & $+63\degr07'48"$ & $16\pm1$               &$32\pm4$    \\
      Ursa Minor     & $15h09m09s$ & $+67\degr13'21"$ & $8.2\pm1.2$            &$76\pm3$    \\
      Willman I      & $10h49m22s$ & $+51\degr03'03"$ & $2.3\pm0.4$            &$38\pm7$    \\
      \hline
     \end{tabular}

    \caption{List of the dwarf spheroidal galaxies used in this analysis, with their J2000.0 coordinates, half-light radius ($r_h$) and distance ($D$).}
    \label{Table-1}
 \end{table}
 
One of the proposed scenarios for indirectly detecting the DM signal is to look for the standard model annihilation/decay products, and the corresponding excess signature in the electromagnetic spectra. A specific case of interest is the weakly interacting massive particles (WIMPs) as DM candidate; here the DM annihilation producing electron-positron pairs through various interaction and decay channels. These charged particles, in turn, can produce broadband electromagnetic emission like bremsstrahlung, inverse Compton, and, particularly at low radio frequency range, synchrotron emission in presence of the, albeit weak, galactic magnetic field \citet{colafrancesco2007, stephano}. Although this signal is expected to be very faint, as the dSph galaxies otherwise have very little or no diffuse low radio frequency emission, due to their expected high DM fraction and their relative proximity, it is in principle possible to detect (or put stringent limit on) this signal, and in turn constrain or rule out certain models on the nature of DM from such observations. Another promising avenue is to look for an excess $\gamma$-ray signal arising due to DM annihilation. Several recent studies have explored this avenue and find some important clues on the nature of DM particle and annihilation cross section \citep{charbonnier2011, walker2011, ackermann2015, brown2019}.

There have been multiple attempts earlier to detect signature of DM annihilation using radio observations of the dSph galaxies. \citet{spekkens}  analysed radio maps of four nearby dwarf spheroidal galaxies observed by GBT at $1.4 \, {\rm GHz}$. Later, \citet{Regis} did deep radio observations of $6$ local group dSph galaxies at $1.1 - 3.1$ GHz using the Australia Telescope Compact Array (ATCA). However, they did not find evidence of any significant emission from DM annihilation. More recently, \citet{kar2019} and \citet{gleam}  analysed low frequency ($72-231$ MHz) radio images of $14$ and $23$ dSphs respectively observed as a part of the GLEAM survey. Although their results are consistent with a null detection, these observations put an upper limit of $\sim 9$ and $\sim 1.5$ mJy~beam$^{-1}$ (for a $\sim 2.5\arcmin$ beam) respectively on the diffuse synchrotron emission from the dark matter. A similar study by \citet{vollmann2020} using LOFAR observations is also consistent with null detection. The results are mostly limited by the low sensitivity and Galactic foreground contamination. Based on the observed limit, they put some constraints on the magnetic field strength as well as DM parameters, e.g., the halo density profile \citep{navarro,spekkens} and the degree of diffusion. Recently, \citet{arpan} have also used  deep ($58$ mJy~beam$^{-1}$) MWA observation of $\omega$~Centauri at $200$ MHz and best fit values of DM particle mass and annihilation cross-section from $\gamma$-ray observations \citep{brown2019} to rule out significant parts of the magnetic field - diffusion coefficient plane. 

In this paper we present an analysis for $23$ dSph galaxies using the $150$ MHz data from the Giant Metrewave Radio Telescope (GMRT). The unique array configuration and high sensitivity of the GMRT allow identification and subtraction of compact sources and, at the same time, constraining the large scale diffuse emission with $\sim 1\arcmin$ resolution. The paper is organized as follows. The dSph sample and the data used for the analysis are described in \S\ref{section:2}. Next, \S\ref{section:3} outlines the analysis and the results are shown in \S\ref{section:4}. Finally, the conclusions are presented in \S\ref{section:5}.

\section{Galaxy Sample and Data}
\label{section:2}

\begin{figure}
	\includegraphics[width=\columnwidth]{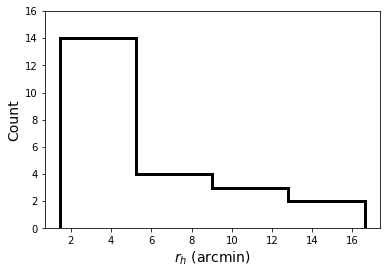}
    \caption{Histogram of the half-light radii, $r_h$ of the sample of 23 dwarf spheroidal galaxies used in this study.}
    \label{fig:figure0}
\end{figure}

Deep, low frequency radio observation of dSph galaxies is sparse, but it is possible to use archival data for a number of targets, and to combine the images of the individual fields to search for weak diffuse emission from the targets \citep[e.g.][]{gleam}. For this work, we have used data from the $150$ MHz TIFR GMRT Sky Survey (TGSS). The first alternative data release (TGSS-ADR1) of this survey \citep[][]{TGSS_ADR1} provides the continuum Stokes I images of $99.5$\% of the sky area north of declination $\delta = -53{\degr}$ (i.e., $\sim 90$\% of the full sky), at a resolution of $25{\arcsec} \times 25{\arcsec}$ for fields that are north of $19{\degr}$ declination, and $25{\arcsec} \times 25{\arcsec}/ \cos(\delta-19{\degr})$ for fields south of $19{\degr}$. The median value of RMS noise is $3.5$ mJy~beam$^{-1}$ for the survey. Out of around 65 dSph galaxies of Milky way and local group \citep[]{McConnachie,dwarf_list}, $23$ are found to be within the TGSS footprint. There were 5 more targets observed in the TGSS, but are not included in this analysis due to presence of image artifacts in those fields. The target coordinates, their half-light radius and distance \citep{McConnachie,gleam,koposov,nicholas,martinez,Laevens_2015} are tabulated in Table~\ref{Table-1}, and the half-light radius distribution is shown in Fig.~\ref{fig:figure0}. Note that the unique hybrid array configuration of the GMRT implies that the telescope has sensitivity to both small scale compact structures as well as large scale diffuse emission, with the largest detectable structure (corresponding to the shortest baselines) being 68 arcmin. If the diffuse emission from these targets have angular extent $\sim$ few $\times r_h$, there will not be any significant missing flux issue in this analysis.

\section{Data Analysis}
\label{section:3}

\begin{figure*}
	\includegraphics[width=82mm]{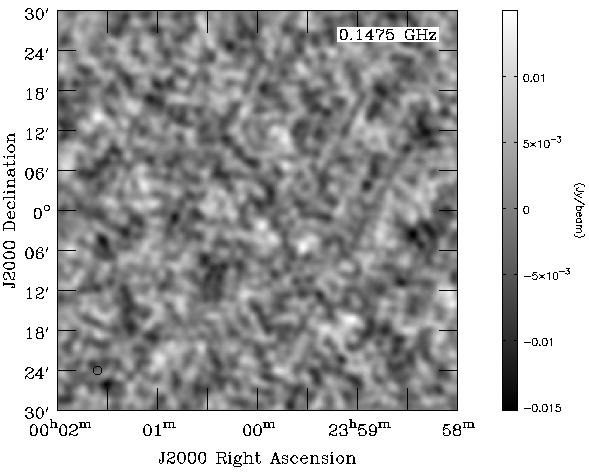} \includegraphics[width=82mm]{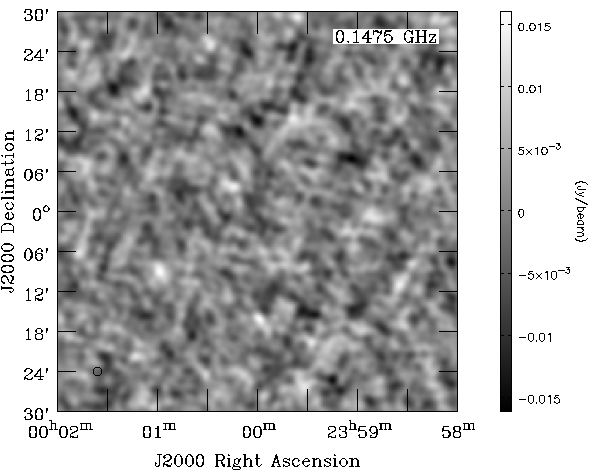}
    \caption{$1\degr \times 1\degr$ stacked image of the target region (left) and the control region (right) using a sample of $23$ dSph galaxies. The angular resolution of the images are $75.6\arcsec$. The colour-bar unit is Jy~beam$^{-1}$.}
    \label{fig:figure1}
\end{figure*}

For the $23$ dSph galaxies, we have used the TGSS-ADR image archive to extract the image of a region centred around the optical position of each of the galaxies in FITS format combining data from multiple overlapping TGSS pointing. For the first part of our analysis, we have extracted $1\degr \times 1\degr$ region for each of the target dSph galaxies. Further analysis was done using the Common Astronomy Software Applications (CASA) package. We have visually inspected each field to identify and blank all the compact sources (unresolved and marginally resolved) using a $5\sigma$ cut off. These compact sources are unrelated to the expected faint diffuse emission (originating from processes directly involving the DM annihilation/decay) at the galaxy-wide scale, that is of our interest. This will also filter out any plausible unidentified imaging artifact present in the field. We then shift the central coordinate of all the fields, regrid and convolve the data to exactly same pixel size and same synthesized beam before combining to create a stacked image from all $23$ dSph target field. We have also selected $23$ ``control'' regions, located close to the corresponding target field regions, that are devoid of any prominent sources. The same process is repeated for these regions to create a stacked image of the control field. We have then compared the properties of the stacked target and control field to search for the presence of any weak, extended emission signal (see \S\ref{section:4}).
 
For the second part of our analysis, instead of stacking a fixed angular size of $1\degr \times 1\degr$, we have used the half-light angular radius ($r_h$) of the dSph galaxies to scale the individual regions before stacking. For this, we start by extracting a subimage of $3r_h \times 3r_h$ for each source. The same steps of removing the compact sources, scaling, regridding and convolving to the same synthesized beam were followed to create the stacked image from the target regions. As earlier, exactly same method was followed to create a stacked image of the control field as well. The results of a careful comparison between the stacked target and control field are presented in the following section.
 
\section{Results and Discussion}
\label{section:4}

\subsection{Comparison between target and control regions}

\begin{figure*}
	\includegraphics[width=82mm]{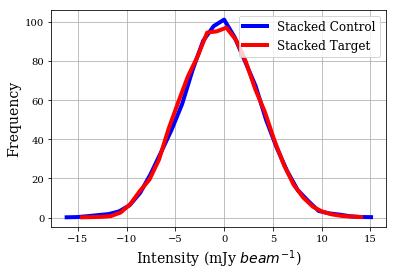} \includegraphics[width=82mm]{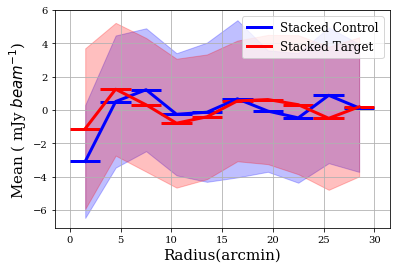} 
    \caption{Left: Normalised intensity distribution for stacked image of the target and the control fields. Right: Radial intensity profiles and $1\sigma$ errors (shaded region) of the stacked target and the control regions.}
    \label{fig:figure3}
\end{figure*}

Fig.~\ref{fig:figure1} shows the final $1\degr \times 1\degr$ stacked target (left) and control (right) regions obtained using the first method described in section \ref{section:3}. As described earlier, the stacking is done after removing all prominent compact sources from individual fields. Even after stacking, no significant large scale emission is visually prominent in either the target field or the control field. The final RMS noise is $\sim4$ mJy~beam$^{-1}$ (with a beam size of $75.6\arcsec$) for both target and control regions. For comparison, the stacked image for the GLEAM survey had a synthesized beam of size $2.28\arcmin$ with the RMS sensitivity 5.5 mJy~beam$^{-1}$. We note that the mean values (0.11 mJy~beam$^{-1}$ for target field and 0.13 mJy~beam$^{-1}$ for control field) are consistent with each other within the error values. A more quantitative comparison between the intensity distributions of the target and control field, shown in Fig.~\ref{fig:figure3} left panel, is also done. We have used the nonparametric Kolmogorov-Smirnov (K-S) test with the pixel intensity values (after regridding to $37$ arcsec); based on the K-S $D$ statistics values ($D = 0.016$ and the $p$-value $0.16$), we conclude that the stacked target and control images intensity distributions are likely to be very similar. The radial intensity profiles (mean surface brightness in different radial bins, and measured RMS in each bin) shown in the right panel of Fig.~\ref{fig:figure3} shows the two stacked images to also have indistinguishable profiles within the statistical uncertainties. 

\begin{figure}
 \includegraphics[width=82mm]{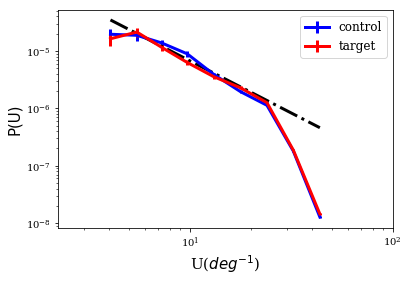}
    \caption{Angular power spectra of intensity fluctuations in the stacked target and the control regions. The dotted black line shows the best fit power law with a power law index of $-1.82$.}
    \label{fig:figure3a}
\end{figure}

\begin{figure*}
	\includegraphics[width=85mm]{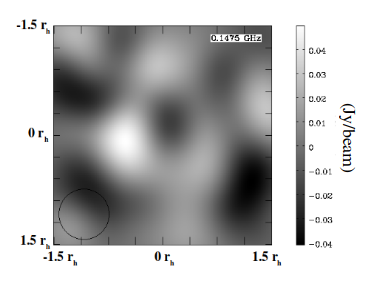} \includegraphics[width=85mm]{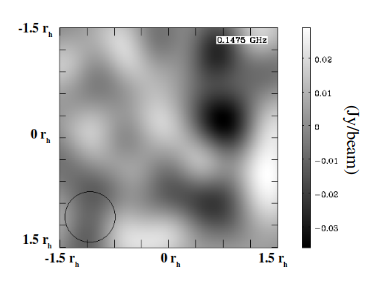}
	\includegraphics[width=85mm]{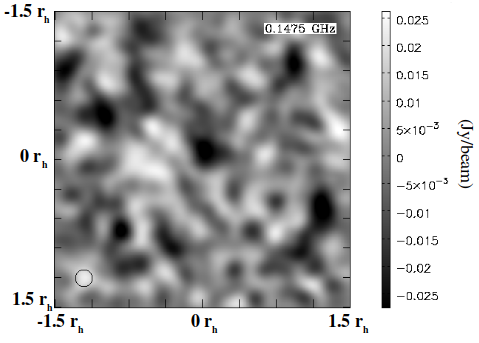} \includegraphics[width=85mm]{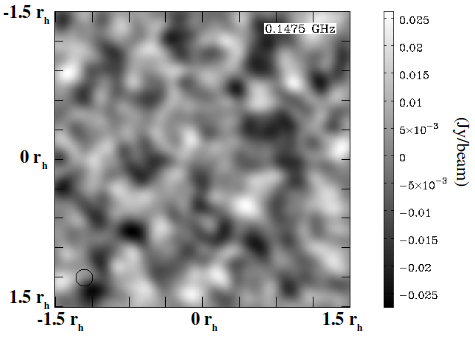}
    \caption{Top: Stacked images of the target (left) and the control region (right) for a sample of $23$ dSph galaxies. Bottom: Same for a subsample of $11$ galaxies with $r_h > 4\arcmin$. The grey scale unit is Jy~beam$^{-1}$, where the beam size is $0.69$ and $0.17$ in $r_h$ unit for the top and the bottom panels, respectively.}
    \label{fig:figure5}
\end{figure*}

Apart from the mean and the distribution function of the intensity values, we have also compared the second order statistics by estimating the angular power spectra of the intensity fluctuations. We convert the image in the Fourier domain using the Fastest Fourier Transform in the West (FFTW; \citealt{frigo05}) and calculate the angular power spectrum. We also multiply the image with a cosine window function to avoid artifacts arising due to the sharp cut-off at the edge \citep{miville16,samir18}. As shown in Fig.~\ref{fig:figure3a}, the angular power spectrum of the stacked target and the control regions are same within the measurement uncertainty. We note that over the $U$ range of $5$ to $30$ {\rm deg$^{-1}$} (i.e., $\sim 2\arcmin - 12\arcmin$), the power spectra can be well represented by a power law with a power law index of $-1.82\pm0.05$ and $-2.05\pm0.09$ for the target and the control region respectively. We choose this $U$ range because $U<5 {\rm deg^{-1}}$ the power spectrum is convolved with the Fourier transform of the window function, and $k>30 {\rm deg^{-1}}$ the image is convolved with the synthesized beam. These measured slopes of the angular power spectrum are consistent with earlier power spectra measurements of the diffuse Galactic synchrotron emission \citep{bernardi09,ghosh12,iacobelli13,samir17,Chakraborty_2019}. Apart from an overall consistency check, this result clearly indicates that the TGSS data are suitable to probe diffuse emission at large angular scale, similar to the expected radio signal associated with the DM halo of the dSph galaxies, present in the field. 

Next, we consider results from a novel stacking method after scaling the radio images of individual dSph fields by the respective half-light angular radius $r_h$. Here, we have assumed that the large scale diffuse signal that we are searching for has characteristic angular scale related to the optical half-light radius ($r_h$). In absence of any direct estimation of the DM halo size of the sample galaxies, this is a reasonable assumption. If that is indeed the case, then, by scaling all the images using the respective $r_h$ and stacking the images, the expected radio emission profiles align with each other and the stacked signal is more likely to be detected. After following the steps described earlier in \S\ref{section:3}, i.e. blanking compact sources, scaling and regridding the images, convolving to a common synthesized beam and stacking, the final images for the target and the control field are shown in the top panels of Fig.~\ref{fig:figure5}. For the two stacked images, the mean values are $0.57$ mJy~beam$^{-1}$ and $0.84$ mJy~beam$^{-1}$ and the rms values are $16$ mJy~beam$^{-1}$ and $11$ mJy~beam$^{-1}$, with the final convolving beam size of $0.69$ in $r_h$ unit, for the target and the control fields respectively.

\begin{figure*}
	\includegraphics[width=82mm]{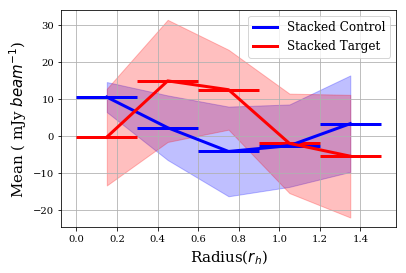}
	\includegraphics[width=82mm]{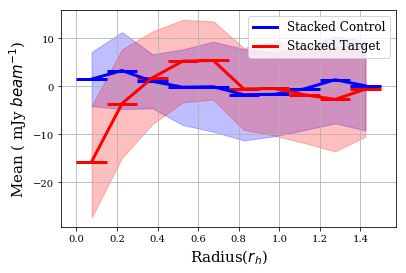}
    \caption{Radial intensity profile for the stacked target and control region when scaled by half-light radius, $r_h$ for a sample of $23$ dSph galaxies (left), for a subsample of $11$ galaxies with $r_h > 4\arcmin$ (right).}
    \label{fig:figure7}
\end{figure*}

One limitation of the analysis with $r_h$ scaling is that the final stacked images have poor resolution. The images are convolved to the same Gaussian beam before stacking, and the resolution is determined by the target with smallest $r_h$ in angular size. We hence also repeat the same analysis for a subsample of $11$ targets with $r_h > 4\arcmin$. The final stacked images of the target and the control fields are shown in the bottom panels of Fig.~\ref{fig:figure5}. For a final convolving beam of $0.17$ in $r_h$ unit, the mean values are 0.9 mJy~beam$^{-1}$ and 0.5 mJy~beam$^{-1}$ where as the rms values are 10 mJy~beam$^{-1}$ and 9 mJy~beam$^{-1}$, respectively.

\begin{figure*}
    \begin{subfigure}{.3\linewidth}
        \centering
        \includegraphics[width=105mm,height=70mm]{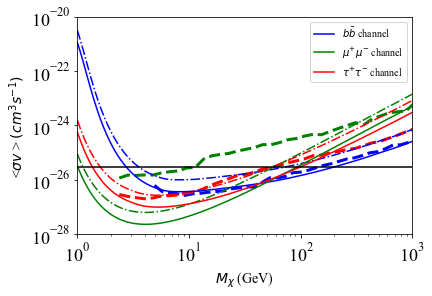}
    \end{subfigure}
    \hfill    
    \begin{subfigure}{.45\linewidth}
        \centering
        \includegraphics[width=65mm,height=40mm]{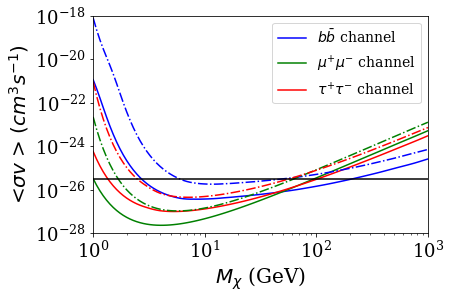}
        \includegraphics[width=65mm,height=40mm]{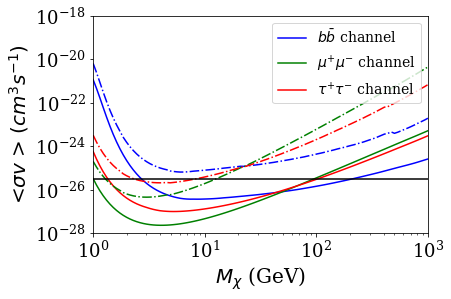}
    \end{subfigure}
    \hfill    
	\caption{Constraints on DM particle mass ($M_{\chi}$) and cross-section(<$\sigma v$>) for three different channels - $b\Bar{b}$ (blue), $\mu^{+}\mu^{-}$ (green) and $\tau^{+}\tau^{-}$ (red). Solid lines in all three panels are for the $5\sigma$ flux density limit from the ``$r_{h}$ stacking'', $B=2~\mu$G and $D_{0}=3\times10^{26} {\rm cm}^{2}{\rm s}^{-1}$. The dot-dash lines are using $5\sigma$ flux density limit from the $1\degr \times 1\degr$ stacking (left panel), $B = 1~\mu$ G (top right panel), and $D_{0} = 3 \times 10^{28} {\rm cm}^{2}{\rm s}^{-1}$ (bottom right panel). The thermal relic cross-section value is marked by solid black line.The current best limits from Fermi-LAT gamma ray observations \citep{hoof2020} for the corresponding decay channels are also shown in dashed lines in the left panel for comparison.}
    \label{fig:figure11}
\end{figure*}

For the ``$r_h$-stacking'' also, we have carried out the similar comparison of the intensity distributions, the radial intensity profiles and the angular power spectra for both the subsample and the full sample. The target and the control region have statistical properties consistent with each other within the measurement uncertainties ($D  = 0.095$ and $p = 0.44$ for the full sample and $D = 0.032$ and $p = 0.13$ for the sub-sample in K-S test, indicating very similar distribution of intensities for the stacked target and control image). As shown in Fig.~\ref{fig:figure7}, there is indication of possible slight excess in the radial profile, however significantly higher sensitivity data are required to check if the feature is real. Overall, the results are consistent with no detection of diffuse large scale signal, and we put some stringent constraints on the galaxy scale diffuse emission at low radio frequency from the dSph galaxies.

\subsection{Constraints on dark matter}

Based on the non-detection of diffuse low frequency emission in these stacking exercises, one can put constraints on DM parameters under reasonable assumptions for the astrophysical conditions in dSph galaxies. This is done by first evaluating the equilibrium electron positron spectra considering diffusion equation (where the source term depends on the DM particle cross-section and mass), and then computing the expected signal for the frequency range of interest due to different emission mechanism \citep[see, e.g.,][for more details]{natarajan,cola,storm,colafrancesco2007,beck,McDaniel,alex,arpan}. A comparison of the expected and observed signal (or the upper limit in case of non-detection) can then constrain the DM particle mass and cross-section. 

For the purpose of deriving the constraints from these observations, publicly available package {\small RX-DMFIT}  \citep{McDaniel} has been used. We choose a very conservative $5\sigma$ upper limit of the estimated flux density at $147.5$ GHz for these computations. The signal is computed for an archetypal dSph galaxy for our sample (similar to the Draco dwarf spheroidal galaxy with core radius $0.22$ kpc and radius for diffusion effect $2.5$ kpc) placed at a distance of $129.5$ kpc, which the median distance for our sample. We also choose $B = 2~\mu$G, and diffusion coefficient $D_{0} = 3 \times 10^{26} {\rm cm}^{2}{\rm s}^{-1}$ \citep{kar2019,natarajan2015green,natarajan,Jeltema_2008,Regis,Chy_y_2011}  as fiducial parameters. As shown in Fig.~\ref{fig:figure11}, we get the best constraint on the average cross-section (<$\sigma v$>) for different DM mass ($M_{\chi}$) for the $5\sigma$ flux density limit of $0.17$ Jy from the ``$r_h$ stacking'' (solid lines in all three panels). The constraints are shown for three different channels, namely $b\Bar{b}$, $\mu^{+}\mu^{-}$ and $\tau^{+}\tau^{-}$ \citep[see][for details]{McDaniel}. The cross-section values are shown for a WIMP mass range of $1 - 1000$ GeV. We note that the constraint on <$\sigma v$> is below the thermal relic cross-section limit \citep{Steigman_2012} for a range of $M_{\chi}$ of interest. 

For comparison, we have also checked how the constraints vary for different choices of these parameters. If we use the $5\sigma$ flux density limit of $0.47$ Jy from stacking of the $1\degr \times 1\degr$ region, the constraints are slightly less tight. This is shown is the left panel of Fig.~\ref{fig:figure11}. The $5\sigma$ flux density limit for the $r_h$ sub-sample stacking is very similar to this value, hence it is not shown separately. Similarly, the top right panel of Fig.~\ref{fig:figure11} shows a comparison between $B = 1$ and $2~\mu$G, and the bottom right panel shows how the constraints change if we assume $D_{0} = 3 \times 10^{28}$ instead of $3 \times 10^{26} {\rm cm}^{2}{\rm s}^{-1}$. The ``no spacial diffusion'' scenario results in constraints very similar to the ones for $D_0 = 3 \times 10^{26} {\rm cm}^{2}{\rm s}^{-1}$. We also show the combined limits from the 11 years Fermi-LAT observations of 27 dSph galaxies \citep{hoof2020} using data available on Zenodo \citep{hoof_data}. The limits derived from the present analysis is comparable to that from the Fermi-LAT observations for $b\Bar{b}$ and $\tau^{+}\tau^{-}$ decay channels, and is more stringent for the $\mu^{+}\mu^{-}$ decay channel. Overall, in all these cases, due to the stringent flux density limits, our results put comparable limits as some of the recent works \citep[e.g.][]{Archambault,Biswas,pooja,vollmann2020} from other radio as well as high energy observations of the dSph galaxies.

\section{Conclusions}
\label{section:5}

Detection of diffuse radio emission from dSph galaxies can be a smoking gun evidence for WIMP as possible dark matter candidate. In this paper, we present a stacking analysis for $23$ dSph galaxies using the archival $150$ MHz data from the TGSS, in order to search for such signal. Quantitative comparisons of first order and second order statistics of the stacked target and control fields show no excess signal for the targets compared to the control regions. Apart from straightforward stacking, we have also tried a novel method of scaling the radio images by the corresponding $r_h$ of the respective dSph galaxies followed by stacking to get a deeper limit on the average flux density from this sample. However, over a range of scale probed in this analysis $U \sim 5 - 30~{\rm deg}^{-1}$, the angular power spectra are found to be power law, with the best fit power law index being consistent with earlier measurements of the diffuse Galactic synchrotron emission. Based on the non-detection of any detectable excess radio emission, we have put tight limits on the thermally averaged cross-section for WIMP mass range of $1 - 1000$ GeV under the assumption of some reasonable astrophysical conditions. With our choices of the parameters, the cross section limit over a significant fraction of the DM mass range is below the thermal relic cross-section limit for $b\Bar{b}$, $\mu^{+}\mu^{-}$ and $\tau^{+}\tau^{-}$ channels. It should be noted that the current analysis is limited by the sensitivity due to the small on-source time per field for the TGSS images. Hence, application of the this method of ``$r_h$ stacking'' with deeper images of the targets from future observations can further improve the limits. Also, combining such observations for larger sample and over the electromagnetic spectrum to put joint limits will also push the constraints. Finally, upcoming telescopes like the Square Kilometre Array (SKA) will have the sensitivity to detect this radio signal or to put orders of magnitude better limits.  

\section*{Acknowledgements}

This work makes use of the NASA/IPAC Extragalactic Database (NED), The SAO/NASA Astrophysics Data System (ADS), and the archival images from the TGSS-ADR1. We thank the staff of the GMRT that made these observations possible. GMRT is run by the National Centre for Radio Astrophysics of the Tata Institute of Fundamental Research. We also acknowledge Alex McDaniel (UCSC) for useful discussion and help with RX-DMFIT. We thank the reviewer for encouraging comments.  We thank Sebastian Hoof for useful discussion on the Fermi-LAT results.

\section*{Data Availability}

The data used for this study is available from the TGSS ADR image archive (data release 1). The final data products from this study will be shared on reasonable request to the corresponding author.






\bsp	
\label{lastpage}
\end{document}